# Maximal Speed of Glucose Change Significantly Distinguishes Prediabetes from Diabetes

## Authors

Dandan Wang[1#], Xiaoyan Chen[2#], Jingxiang Lin[2], Teng Zhang[1], Lianyi Huang[2], Dongliang Leng[1], Xiaohua Douglas Zhang[1*], Gang Li[1*]

## Affiliations

1. Department of Public Health and Medicinal Administration, Faculty of Health Sciences, University of Macau, Taipa, Macau
2. Department of Endocrinology, the First Affiliated Hospital of Guangzhou Medical University, Guangdong, China

[#]: Wang D and Chen X contributed equally to the work.

## * Correspondence:

Gang Li, PHD, Faculty of Health Sciences, University of Macau, Taipa, Macau. Email: gangli@um.edu.mo

Xiaohua Douglas Zhang, PHD, Faculty of Health Sciences, University of Macau, Taipa, Macau. Email: douglaszhang@um.edu.mo. Current affliliation: Department of Biostatistics, University of Kentucky, Lexington, KY 40536, USA.

**Abstract**

*Introduction*: Rapid changes in blood glucose levels can have severe and immediate health consequences, leading to the need to develop indices for assessing these rapid changes based on continuous glucose monitoring (CGM) data.

*Research design and methods*: We proposed a CGM index, maxSpeed, that represents the maximum of speed of glucose change (SGC) in a subject, respectively, and conducted a clinical study to investigate this index along with SGC mean (meanSpeed) and SGC standard deviation (sdSpeed), coefficient of variation (CV), standard deviation (SD), glycemic variability percentage (GVP), mean amplitude of glycemic excursions (MAG), mean absolute glucose excursion (MAGE), mean of daily differences (MODD) and continuous overlapping net glycemic action (CONGA).

*Results*: Our study revealed that, there exist multiple patterns in distinguishing non-diabetes, prediabetes, type 1 diabetes (T1D) and type 2 diabetes (T2D). First, maxSpeed significantly distinguishes between either of non-diabetes and prediabetes and either of T1D and T2D. Second, meanSpeed, sdSpeed, GVP and MAG significantly distinguish between non-diabetes and either of T1D and T2D. Third, MODD and CONGA of 24 hours significantly distinguish between non-diabetes and either of T1D and T2D, between T1D and either of prediabetes and T2D. Fourth, SD, MAGE and CONGA of 12 hours significantly distinguish between non-diabetes and either of T1D and T2D, between T1D and pre-diabetes. Fifth, CV significantly distinguishes between T1D and either of Non-diabetes and T2D.

*Conclusion*: The CGM index, maxSpeed, is easy to calculate and readily understandable to clinical scientists. It significantly distinguishes prediabetes from T2D as no other indices under investigation have such an ability. maxSpeed assesses the rapid change of glucose in a short term, which is important both biologically and clinicially because our human body may not tolerate too rapid change in a short term.

**Introduction**

Diabetes is one of the world's most prevalent diseases. From the World Health Organization, an estimated 8.5% (i.e., 422 million) of adults aged 18 years and older globally were living with diabetes.[1] In order to manage diabetes and help avoid its associated problems, patients with diabetes need to monitor their blood glucose levels closely. To facilitate glucose monitoring, continuous glucose monitoring (CGM) devices were introduced to diabetes care.[2,3] The CGM device uses a sensor inserted under the skin to detect glucose levels in tissue fluid so as to continuously measure the glucose levels in tissue fluid over a relatively long period of time, such as 7 or 14 days. These devices are useful in detecting fluctuation and trends of blood glucose levels which are important for disease diagnosis and treatment.

Various CGM indices have proposed. Early in 1970, Service et al. proposed the mean amplitude of glycemic excursions (MAGE),[4] which is the mean of blood glucose values exceeding one standard deviation from the 24-hour mean blood glucose. Mean absolute glucose (MAG) which is computed by calculating the sum of absolute differences between successive glucose values and then divided by the total time.[5] MODD, the mean of daily differences, estimates the between-day glycemic variability (GV) and it is defined as the mean of the absolute difference between two glucose readings tested at the same time within a 24 interval.[6,7] Continuous overlapping net glycemic action (CONGA) is another metric to measure GV by integrating the duration and degree of glucose excursions.[8] A *n*-hour CONGA is the standard deviation of the difference between two glucose values being n hours apart. Peyser et al proposed the glycemic variability percentage (GVP) which is the length of the continuous glucose monitoring (CGM) temporal trace normalized to the duration under evaluation.[9] Some of these metrics are correlated[10,11] and all of them focus on the overall variability of glucose fluctuation. None of them directly assess rapid swing in blood glucose. Rapid swings in blood glucose levels can have severe and immediate health consequences. Here we proposed a new metric, maximal speed of glucose change (maxSpeed), to capture this rapid swing in blood glucose levels based on CGM data. We also explored two more metrics based on the speed of glucose change (SGC), mean of SGC (meanSpeed) and standard deviation of SGC (sdSpeed) to capture the overall feature of SGC for a subject.

Prediabetes is also called impaired glucose tolerance or intermediate hyperglycemia in which the glycemic parameters of prediabetes are above normal but below diabetes thresholds. It is vital to conduct research in refining and developing diagnostic criteria to accurately identify individuals with prediabetes and differentiate them from those with diabetes because these criteria are essential for early detection, intervention, and prevention strategies for diabetes. Here, we conducted a CGM clinical study including non-diabetes, prediabetics, T1D and T2D (Clinical trial registration number: ChiCTR-OOC-17010974). Using this study we investigated the utility of maxSpeed, meanSpeed and sdSpeed in distinguishing prediabetes from diabetes and compare them with other CGM indices such as standard deviation of glucose values (SD), coefficient of variability of glucose values (CV), GVP, MAG, MAGE, CONGA based on CGM data. This study shows that MaxSpeed can significantly distinguish predicates from not only T1DM but also T2DM. The results of this study may not only provide reference for research on the dynamics of glucose fluctuation but also open a new avenue for the management and care of diabetes, especially with regards to precision medicine for diabetes.

**Research Design and Methods**

Based on the regular tests for diabetes and the judgment of medical doctors in the hospital, a total of 125 individuals, including 24 T1D and 67 T2D, 9 prediabetes and 25 healthy people, were enrolled in our study. All T1D or T2D patients were treated with insulin and/or oral anti-diabetic agents. Participants were requested to eat their usual diet at usual amount and time, and to avoid overeating and excessive exercise during monitoring phase. Informed consent was obtained from all participants. The study was conducted according to the principles of the Helsinki declaration. It was designed that the subcutaneous interstitial fluid glucose data were obtained every 3 minutes for 7 days using the Glutalor's CGMs device (European Registration number: QMF-MF-33013SHG, China Registration number: CQZ1600116) for all participants. Participants were required to measure fasting glucose value every morning with the blood glucose meter for calibration.

To ensure data quality, we excluded the CGM data of participants with the number of non-missing glucose values less than 720 (i.e., with less than one and half a day of non-missing values). If a patient used more than one device, we only used the one with the most completed data for further analysis. The CGM data of the three subjects excluded for further analysis by the quality control process are shown in Panels 1 to 3 of Supplemental Figure 1. The demographics of the participating subjects after quality control in this clinical study is shown in Supplementary Table 1.

In this research, we investigated three GV indices based on speed of glucose change to measure the fluctuation of glucose dynamics as follows. Speed of glucose change (SGC) at time $t_i$ is defined as the ratio of the amount of glucose change (i.e., $|\Delta y_i|$) to the length of time changed (i.e., $|\Delta t_i|$) in a short period of time (Figure 1). That is, the SGC at time $t_i$ is $\frac{|\Delta y_i|}{|\Delta t_i|}$ (Figure 1B). Normally, SGC has the unit of mg/dL/minute or mmol/L/minute. For a subject, there will be many SGC values, each corresponding to one measured time point. The maximum of these SGC values for a subject is defined as maxSpeed. Likewise, the mean and standard deviation of these SGC for a subject are denoted as meanSpeed and sdSpeed, respectively. Note, sdSpeed is also called standard deviation of blood glucose rate of change.[12]

We calculated maxSpeed, meanSpeed and sdSpeed along with multiple other CGM indices such as SD, CV, GVP, MAG, MAGE, MODD, CONGA based on the CGM data of each subject. For each of these indices, we will use unpaired t-tests to test the difference between any pair of groups among the four groups of subjects (i.e., non-diabetics, prediabetes, T1D and T2D) and adjust the p-value using Benjamin-Hochberg correction on multiplicity.[13] The p-value adjusted using Benjamin-Hochberg correction on multiplicity is denoted as BH adjusted p-value here. Because the variance in the four groups is not all equal, the pairwise t-test is conducted. Statistical analysis was conducted using R.

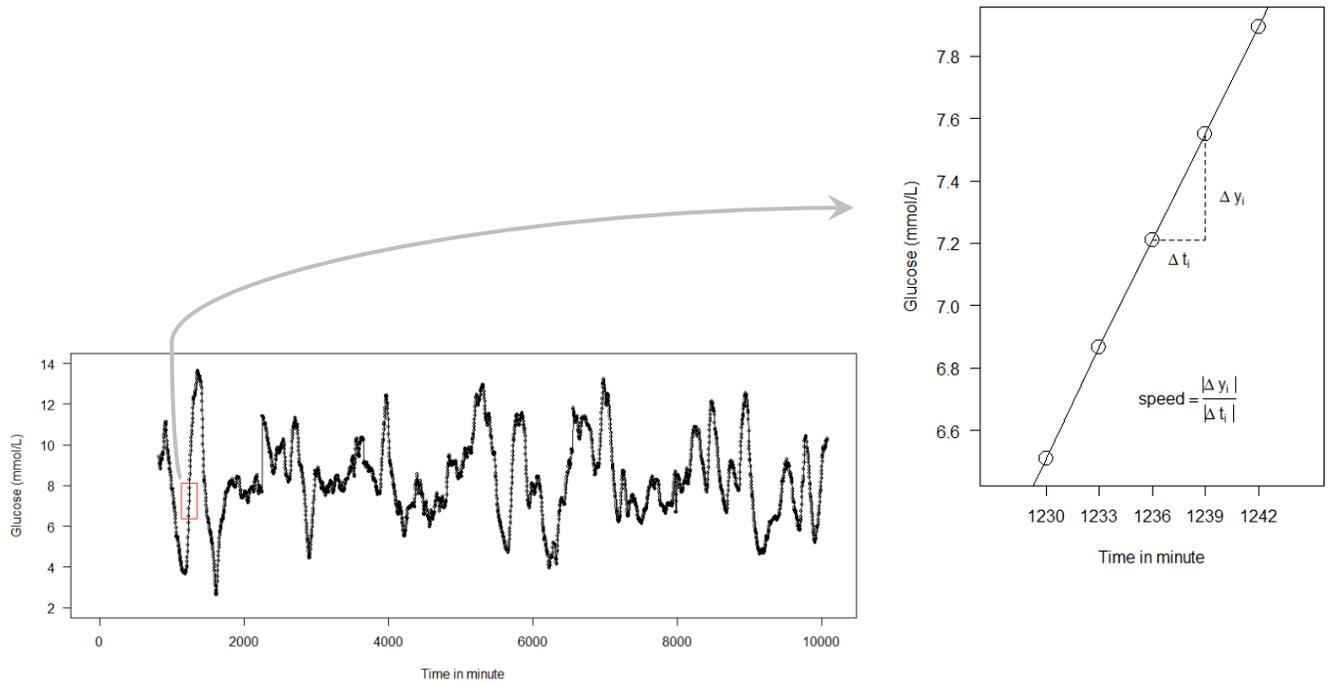

**FIG. 1.** Speed of glucose change to assess the fluctuation in the glucose dynamics measured by the CGM device. A: glucose dynamics of a diabetes patient. B: definition of speed of glucose dynamics illustrated by a segment in the red box in Panel A.

## Results

After obtaining maxSpeed, meanSpeed, sdSpeed, SD, CV, GVP, MAG, MAGE, MODD and CONGA for each subject, the mean and standard error of these indices in the subjects of each of the four groups (i.e., non-diabetes, prediabetes, T1D and T2D) were calculated and are shown in Figure 2. The significant results (i.e., BH adjusted p-value being less than 0.05) of pairwise comparisons are also shown in Figure 2. The results displayed in Figure 2 can be summarized as follows.

First, maxSpeed significantly distinguishes between T1D and non-diabetes (BH-adjusted p-value=0.0376), between T2D and non-diabetes (adjusted p-value=0.0103), between T1D and prediabetes (adjusted p-value=0.0376), between T2D and prediabetes (adjusted p-value=0.0376), but cannot significantly distinguish between prediabetes and non-diabetes (adjusted p-value=0.8236) or between T1D and T2D (adjusted p-value=0.4424).

Second, meanSpeed only significantly distinguishes between T1D and non-diabetes and between T2D and non-diabetes, but not other pairs. So are sdSpeed, MAG and GVP.

Third, SD significantly distinguishes between T1D and non-diabetes, between T2D and non-diabetes and between T1D and pre-diabetes, but not other pairs. So are MAGE and 12-hour CONGA (i.e., CONGA.12hr).

Fourth, MODD and 24-hour CONGA (i.e., CONGA.24hr) shares another pattern of significant results in distinguishing disease status of diabetes. That is, they significantly distinguish between T1D and non-diabetes, between T2D and non-diabetes, between T1D and prediabetes and between T1D and T2D, but neither between prediabetes and non-diabetes nor between prediabetes and T2D.

Fifth, the mean of glucose values significantly distinguishes between T1D and non-diabetes, between T2D and non-diabetes and between T2D and pre-diabetes, but not other pairs.

Sixth, CV only significantly distinguishes between T1D and non-diabetes, between T1D and T2D, but not other pairs.

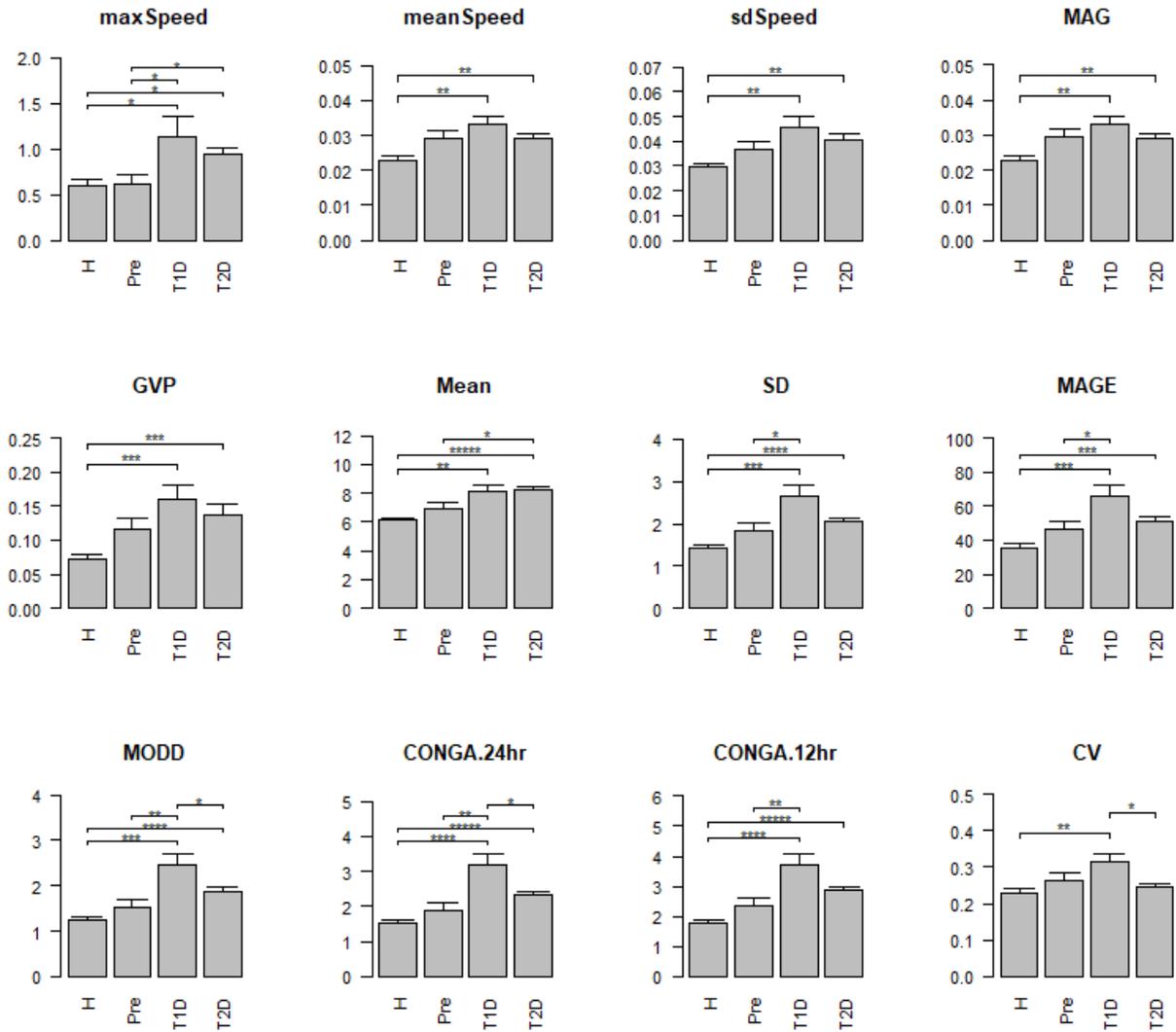

**FIG. 2.** Bar-error plots to show pairwise comparison among groups of subjects, i.e., nondiabetic (H), prediabetes (Pre), type 1 diabetes (T1D) and type 2 diabetes (T2D) of different metrics on glycemic variability. The height of the bar represents the mean and the length of the vertical segment above a bar represents standard error. "*", "**", "***", "****" denote the adjusted p-value less than 0.05, between 0.05 and 0.01, between 0.01 and 0.001, and between 0.001 and 0.0001, respectively.

## Discussion

For the use of CGM for diabetes care, it is critical to investigate the fluctuation of glucose dynamics concealed in the CGM data.[6,14] Recently review articles[6,12,15-17] described various CGM indices including SD, CV, GVP, MAG, MAGE, MODD and CONGA. Each index may provide insights into different aspects of CGM data and the developments of indices for better defining and deciphering GV based on CGM data should help to improve understanding of the clinical relevance of GV in the management of diabetes[6]. Here we also conducted a clinical study to investigate maxSpeed, meanSpeed, sdSpeed, SD, CV, GVP, MAG, MAGE, MODD and CONGA, with a special attention to the index capture the speed of rapid change in a short term. Our study revealed that, there exist multiple patterns in the ability of distinguishing non-diabetes. prediabetes, T1D and T2D, as shown in Figure 2, which is described in the following paragraphs.

The meanSpeed, sdSpeed, GVP and MAG share a pattern of significantly distinguishing between T1D and non-diabetes and between T2D and non-diabetes but no other pairs of comparisons. It is not surprising that the patterns of meanSpeed, MAG and GVP are all the same because, as shown the in the Method Section, meanSpeed equals MAG and provides a low bound for GVP when the interval $|\Delta t_i|$ between any two successive points is equal. It might be coincident that the pattern of meanSpeed also equals that of sdSpeed.

SD and MAGE share a pattern of significantly distinguishing between T1D and non-diabetes, between T2D and non-diabetes and between T1D and pre-diabetes, but not other pairs. This result is consistent with the high correlation between SD and MAGE (i.e., ) shown by Rodbard[10]. The 12-hour CONGA also shares the same pattern as SD and MAGE in this study, which is worthwhile to explore whether this always happens in the future research.

MODD and 24-hour CONGA share another pattern of significantly distinguishing between T1D and non-diabetes, between T2D and non-diabetes, between T1D and prediabetes and between T1D and T2D, but neither between prediabetes and non-diabetes nor between prediabetes and T2D. The observation that MODD and 24-hour share the same pattern is consistent with the simple linear relationship between MODD and 24-hour CONGA (i.e., CONGA.24hr=1.22MODD) under the assumption of a normal distribution for the glucose values provided by Rodbard[11]. In this study, SD did not show the same pattern as, but only similar to, that of MODD and 24-hour CONGA although Rodbard[11] also provides a linear relationship between MODD and SD (i.e., MODD = 1.16 SD) and between CONGA and SD (i.e., CONGA = 1.42 SD). This may be explained by the fact that, beside the assumption of a normal distribution for the glucose values as being needed to derive the relationship between MODD and 24-hour CONGA, these two linear relationships between SD and MODD and between SD and 24-hour CONGA rely on one more assumption that the difference of glucose values in two time points with 24-hour apart has the same distribution across different days.

CV has a pattern of significantly distinguishing between T1D and Non-diabetes, between T1D and T2D, but not other pairs of comparison. This result is consistent with those revealed by multiple studies, that is, CV was higher in T1D than in T2D.[18,19]

It is remarkable to discover that maxSpeed has a unique pattern in which it significantly distinguishes between T1D and Non-diabetes, between T2D and Non-diabetes, between T1D and prediabetes, between T2D and prediabetes, but does not significantly distinguish between prediabetes and non-diabetes or between T1D and T2D. Even more remarkable is the ability of maxSpeed to significantly distinguish between prediabetes and T2D as no other CGM indices under investigation have such an ability.

SGC is simply the amount of glucose change per minute. Because a person may have different SGC at different time, it is important to investigate the maximum, mean and standard deviation of SGC (represented by maxSpeed, meanSpeed and sdSpeed, respectively) for a person. The concepts of SGC, maxSpeed, meanSpeed and sdSpeed should be readily understandable to clinical practitioners and biomedical researchers. Missing values commonly exist in CGM data[14]. Because ignoring the missing values has little impact on the calculation of maxSpeed, meanSpeed and sdSpeed for CGM data with two or more days, maxSpeed, meanSpeed and sdSpeed are much more robust to missing values as compared to other CGM indices like CONGA and MODD.

MAG, GVP and meanSpeed are directly or indirectly related to SGC on average as shown in the Method section. Thus, they can more or less be explained or illustrated using SGC. In fact, as demonstrated in this clinical study, MAG, GVP, meanSpeed and sdSpeed all have the same ability in distinguishing diabetes disease status. A reminder in the calculation of SGC is that only non-missing glucose values in two consecutive points in the minimal time interval (such as 5 minutes in Dexcom, Medtronic, Free Libre and 3 minutes in Glutalor CGM systems) can be used for the calculation of SGC.

The inherent maximal speed of glucose change reflects rapid short-term fluctuations in blood glucose levels. Studying these rapid changes is important because the human body may not tolerate abrupt shifts in glucose levels over short periods. This consideration underlies the use of maxSpeed as a metric. However, a key caveat is that maxSpeed estimates can be sensitive to outliers resulting from measurement errors. Smoothing the glucose time series may not be suitable in this context, as it could obscure biologically meaningful rapid changes. Therefore, future research is needed to refine the estimation of maxSpeed in a way that reduces sensitivity to noise while preserving genuine physiological signals. Another limitation is that the patients with type 1 and type 2 diabetes in this study were under medical treatment; thus, caution is warranted when applying these CGM-derived findings to screen individuals who are not receiving treatment.

## Acknowledgments

This work was supported by the Science and Technology Development Fund, Macau SAR (File no. 0004/2019/AFJ and 0011/2019/AKP).

## Competing interests

The authors declare that the research was conducted in the absence of any commercial or financial relationships that could be construed as a potential conflict of interest.

# Supplementary Materials

Supplementary Table 1. demographics of the participating subjects in the clinical study

|  | Type 2 D | Type 1 D | Prediabetes | Non-diabetes |
|---|---|---|---|---|
| Number | 64 | 24 | 9 | 25 |
| Gender(M/F) | 43/21 | 12/12 | 5/4 | 13/12 |
| Age, years | 56.64 ± 14.38 | 35.26 ± 17.11 | 48.84 ± 16.93 | 34 ± 7.96 |
| Height, cm | 170.86 ± 8.98 | 164 ± 6.29 | 171 ± 3.82 | 169.34 ± 4.06 |
| Weight, kg | 72.4 1± 8.99 | 54 ± 15.66 | 74.81 ± 15.56 | 71.68 ± 7.67 |
| BMI, kg/m$^2$ | 24.76 ± 8.89 | 21.42 ± 3.48 | 25.63 ± 3.89 | 22.30 ± 3.79 |
| HbA1C, % | 7.87 ± 2.65 | 8.38 ± 2.29 | 5.83 ± 1.56 | 4.96 ± 0.38 |

Note: Data are expressed as means ± SD. BMI, body mass index; HbA1c, glycated hemoglobin; No significant difference in sex, height, weight, and BMI between the diabetic patients and non-diabetic people assessed by Mann-Whitney U-test.

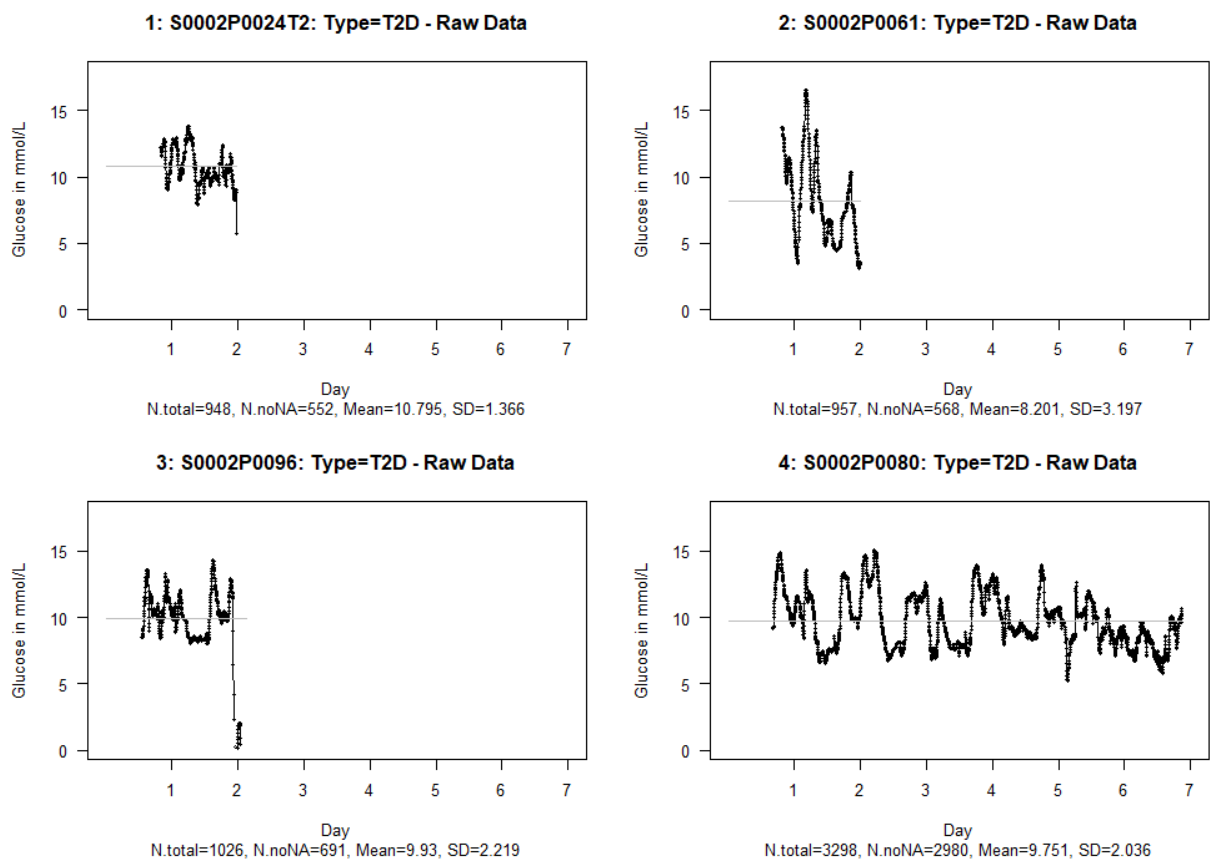

Supplementary Figure 1. Time series plot for glucose values measured by the continuous glucose monitoring device in three subjects failing the quality control (Panels 1 to 3) as contrast to a subject with CGM data passing quality control (Panel 4).